\documentclass[]{raa}
\usepackage{cases}            % referee version: for submission
\usepackage{graphicx,times,subfig}
\usepackage{natbib}
\usepackage{longtable,textcomp}

\begin{document}

   \title{X-ray afterglow of GRB 050712: Multiple energy injections into the external
   shock
%$^*$ \footnotetext{\small $*$ Supported by the National Natural
%Science Foundation of China.}
}

  \volnopage{Vol.0 (200x) No.0, 000--000}
   \setcounter{page}{1}

   \author{Liang-Duan Liu \inst{1,2} and A-Ming Chen\inst{1} }

\institute{ $^1$Institute of Astrophysics, Central China Normal
University, Wuhan 430079, China \\
$^2$School of Astronomy and Space Science, Nanjing University,
Nanjing 210093, China
 }


\abstract{As indicated by the observed X-ray flares, a great
amount of energy could be intermediately released from the postburst
central engine of gamma-ray bursts (GRBs). As a natural consequence,
the GRB external shock could be energized over and over. With such a
multiple energy injection model, we explore the unique X-ray
afterglow light curve of GRB 050712, which exists four apparent
shallow decay plateaus. Together with three early X-ray flares, the
central engine of GRB 050712 is supposed to release energy at least
seven times after the burst. Furthermore£¬ we find that the energy
released during four plateaus are all on the same order of
magnitude, but the luminosity decreases with time significantly.
These results may provide some interesting implications for the GRB
central engine.
 }

    \authorrunning{L.-D. Liu \& A.-M. Chen}            %author_head in even pages
    \titlerunning{X-ray afterglow of GRB 050712}  % title_head in odd pages
    \maketitle

%________________________________________________ sections below
%
\section{Introduction}           %% first-level sections will be auto-capitalized
\label{sect:intro} The successful launch and operation of the
\textit{Swift} satellite have significantly improved our
understanding of the physical origin of gamma-ray bursts (GRBs). The
X-ray Telescope (XRT) onboard \textit{Swift} has revealed many
unexpected features in X-ray afterglows (Zhang et al. 2006), which
include: (1) steep decay phase (Zhang et al. 2007; 2009), (2)
shallow decay phase (Liang et al. 2007), (3) missing jet breaks in
some GRBs (Liang et al. 2008; Racusin et al. 2009), and (4) X-ray
flares (Falcone et al. 2007; Swenson et al. 2013). Specifically, the
shallow decay segments were commonly seen in about $60\%$ of the
\textit{Swift}-detected X-ray afterglows, and the flares appear in
roughly $1/3$ of all (Gehrels et al. 2009; Grupe 2013). It seems
difficult to explain these phenomena in the standard external shock
model (M{\'e}sz{\'a}ros et al. 1997; Sari et al. 1998; Zhang 2007;
Liang et al. 2007), where a simple power-law decay with a temporal
index around 1.2 is predicted in the X-ray light curves.

X-ray flares share a lot of similarities with GRB prompt emission,
which indicates that the flares might have similar physical
processes with the prompt emission including the dynamics and
radiation mechanisms (Fan \& Wei 2005; Zhang,B. et al. 2006; Yu \&
Dai 2009). At the same time, it is widely accepted that the shallow
decay afterglows could be due to an energy injection into the
external shock (Dai \& Lu 1998a,b; Rees \& M\'esz\'aros 1998; Zhang
\& M{\'e}sz{\'a}ros 2001; Fan \& Xu 2006; Nousek et al. 2006; Granot
\& Kumar 2006; Sollerman et al. 2007; Geng et al. 2013). Both the
X-ray plateaus and flares suggest that the GRB central engine
(accreting black hole or millisecond magnetar) has not been switched
off immediately after the prompt emission. For a black hole
surrounded by an accretion disk, the late activities could be
powered by the accretion of the fall-back materials, where the spin
energy of the black hole can be released by the Blandford-Znajek
mechanism (Blandford \& Znajek 1977). On the other hand, if the
central engine is a millisecond magnetar, it will lose its
rotational energy by some baking mechanisms (e.g., magnetic dipole
radiation).

Long GRB 050712 with a duration of $T_{90} = 48 \pm 2$ s was
detected by the BAT onboard \textit{Swift} at 14:00:28 UT on 2005
July 12, and the XRT refined coordinates $\rm
RA(J2000)=05^{h}10^{m}48^{s},~ Dec(J2000)=64^{\circ}55^{'}48.2^{''}$
with a $6^{''}$ radius error circle. Its peak flux measured in the
15$-$150 keV band was $(1.10\pm0.07) \times10^{-6}$ erg ${\rm
cm^{-2}~s^{-1}}$. The XRT observation in the 0.3$-$10 keV band
started at about 160 s after the trigger. The observed X-ray light
curve exhibits a unique temporal behavior. Firstly, three flares
happened at 174, 455 and 861s (Swenson \& Roming 2013) after the BAT
trigger respectively. Such early flares are usual in Swift GRBs
(Falcone et al 2007; Zhang,B. et al 2006). However, after these
flares, the afterglow emission declines tierdly (i.e., step by step;
see Figure 1), which is completely different from the normal
single-power-law decay or a single plateau followed by a power-law
decay. Furthermore, the rise and decline of these steps are very
slow, so they can not be late flares. Obviously, such a tierd light
curve can not be explained by the traditional energy injection model
where the energy is continuously injected. Alternatively, in this
paper we suggest that the afterglow emission of GRB 050712 could
indicate several intermediate, multiple energy injections.

%In this paper, we fit the X-ray afterglow of GRB 050712 by using
%several times energy injection model. In Section 2, showing the
%observational facts of GRB 050712. In Section 3, we describe  the
%several times energy injection model in detail, including the
%dynamics and the radiation mechanism. In Section 4, we show the
%numerical result of the model described above, and fit the unusual
%X-ray afterglow light curves of GRB 050712. Finally, in Section 5,
%we summarize our finding and give a brief discussion.

% Authors can give a citation as `\citealt{Michel+etal+1992}'.
% You may also use \cite, \citep and \citet for citation, and use Table~1
% or Figure~1 and so forth. Using \ref and \label for cross-references of
% Tables/Figures is a good way in adjusting/adding/removing text, tables or
% figures.

\section{MODEL}
In the standard model of GRBs (Zhang 2007; Gehrels et al. 2009), the
GRB central engine drives an energetic fireball or
Poynting-flux-dominated ejecta, which could be accelerated to a high
relativistic speed. Then the internal dissipations in the ejecta
(e.g. internal shocks and magnetic reconnections) lead to the prompt
gamma-ray emission. Subsequently, the interaction between the GRB
ejecta and the ambient medium gives rise to a relativistic external
shock, which could produce long-lasting multi-wavelength afterglow
emission. Therefore, the temporal behavior of the afterglow emission
should be basically determined by the dynamical evolution of the
external shock. A generic dynamical model which covers both the
ultra-relativistic and non-relativistic shock dynamics was
delineated by Huang et al.(1999;2000b). The fundamental dynamical
equation can be written as
\begin{equation}
\frac{d\Gamma}{dm_{\rm sw}}=-\frac{\Gamma^{2}-1}{M_{\rm ej}+2\Gamma
m_{\rm sw}},
\end{equation}
where $\Gamma$ is the bulk Lorentz factor of the external shock,
$m_{\rm sw}$ is the mass of the swept-up medium, and $M_{\rm ej}$ is
the mass of the GRB ejecta.

The post-burst energy injection into the external shock could
significantly slow down the deceleration rate of the shock or even
lead to a short-term acceleration (Dai \& Liu 2012). The dynamical
equation can be derived from the energy conservation law. The total
energy of the GRB ejecta and the shocked medium can be written as
$E_k=\Gamma M_{\rm ej}c^2+(\Gamma^2-1)m_{\rm sw}c^2$, where the
shock jump condition is used to calculated the internal energy. Then
we can get $dE_k=(\Gamma^2-1)c^2dm_{\rm sw}+(M_{\rm ej}+2\Gamma
m_{\rm sw})c^2d\Gamma=L\left(t\right)dt$, where $L$ is the
luminosity of the post-burst energy flow released from the central
engine and $t=t_{\rm obs}/(1+z)$ with $t_{\rm obs}$ being the time
measured in the observer's frame and $z$ redshift (Fan \& Xu 2006).
Therefore, we can revise the dynamical equation as follows
\begin{equation}
\frac{d\Gamma}{dm_{\rm sw}}=-\frac{1}{M_{\rm ej}+2\Gamma m_{\rm
sw}}\left[\Gamma^{2}-1-{L\left({t}\right)\over c^2}{dt\over dm_{\rm
sw}}\right].
\end{equation}
If $L=0$ (no energy injection), the above equation returns to Eq.
(1). For a general consideration, Zhang \& M{\'e}sz{\'a}ros (2001)
proposed a simple power-law form for the injected luminosity as
$L(t)\propto t^{-q}$. Specifically, $q=5/3$ is expected if the
central object is a black hole surrounded by a fallback accretion
disk. However, for $q>1$, the energy injection is not important. On
the other hand, for a millisecond magnetar, $q=0$ or $-2$ can be
found for a traditional magnetic dipole radiation before/after the
spin down timescale (Dai \& Lu 1998a). Moreover, for a
fallback-accreting magnetar, its spin down history could become much
more complicated (Dai \& Liu 2012) and then the value of $q$ might
evolve with time.

In most energy injection models, the energy release is usually
assumed to be continuous. However, as indicated by the X-ray flares,
the energy release from the post-burst object could be intermediate.
It seems that the central engine can be turned on after a period of
energy accumulation. Therefore, it is reasonable to consider that
the energy injection into the external shock might be intermediate.
In this case, we have
\begin{equation}
L_i(t)=Q_{i}t^{q_{i}},~~ {\rm for} ~~t_{{\rm strat},i}<t<t_{{\rm
end},i},
\end{equation}
where the subscript ``$i$" represents the $i$th energy injection,
$Q_i$ and $q_i$ are constants in each injection, and the energy
injection starts and ends at $t=t_{{\rm start},i}$ and $t_{{\rm
end},i}$ respectively. For an example, Geng et al. (2013) used a
two-step energy injection model to interpret the rebrightening of
the multi-band afterglow of GRB 081029.
\begin{figure}
 %  \vspace{0.5cm}
  % \begin{center}
  % \plotone{ms1391fig3
%.eps}
  % \begin{minipage}[]{85mm}
\centering
\resizebox{0.7\hsize}{!}{\includegraphics{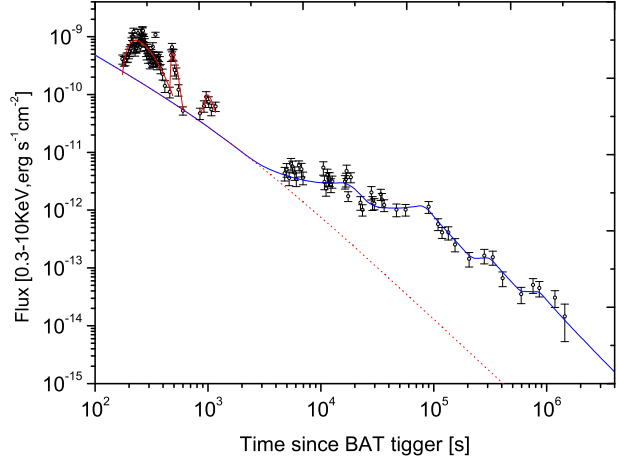}}
   \caption{ Fitting to the X-ray afterglow
   of GRB 050712 with multiple energy injections (solid blue line), where the case without energy injection is also presented for a comparison (dotted line).
   The three early flares are fitted by the empirical ``Norris
   function" function (solid red line).
   The observational data are
   taken from $http://www.swift.ac.uk/xrt$\_$curves/00145581/$. }
 %  \end{center}
   \label{Fig:plot1}
   \end{figure}

\begin{figure}
\centering
\resizebox{0.5\hsize}{!}{\includegraphics{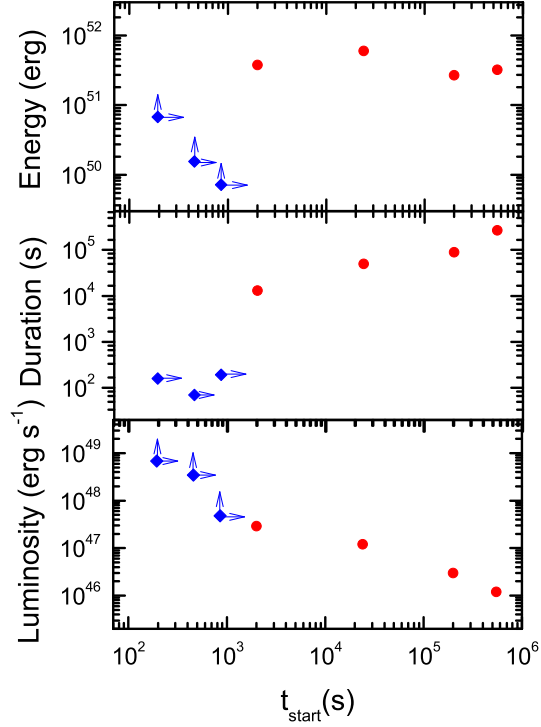}}
\caption{The energy (top), duration (middle), and luminosity
(bottom) of the energy releases by the central engine during three
flares (diamonds) and the four plateaus (circles).}
\end{figure}

In order to solve equation (2), the evolutions of the swept-up mass
$m_{\rm sw}$ and the radius $r$ of the shock should be described
together as (Huang et al. 1999, 2000b)
\begin{equation}
\frac{dm_{\rm sw}}{dr}=2 \pi r^{2}(1-\cos\theta_{j}) n m_{\rm p},
\end{equation}
\begin{equation}
\frac{dr}{dt}={\beta c\over 1-\beta},
\end{equation}
where $n$ is the ambient number density, $\theta_{j}$ the half
opening angle of the jet, $m_{p}$ the mass of proton, and
$\beta=\sqrt{1-\Gamma^{-2}}$ the velocity in the unity of the speed
of light. The distribution of the ambient medium plays an important
role in determining the profile of the afterglow light curve. On
contrary to the usually assumed homogenous interstellar medium
(ISM), a wind-like circumstance with $n\propto r^{-2}$ is also
proposed by Dai \& Lu (1998c) and Chevalier \& Li (2000) by
considering that long GRBs probably originate from the collapse of
massive stars (Woosley 1993; MacFadyen \& Woosley 1999).

As the propagation of the external shock, charge particles are
accelerated and the stochastic magnetic field is amplified. The
internal energy of the shocked medium is shared by magnetic fields
and electrons with fractions $\epsilon_{B}$ and
$\epsilon_{e}\sim\sqrt{\epsilon_{B}}$ (Medvedev 2006), respectively.
The accelerated electrons are further assumed to distribute as
$n_{{\gamma'}}^{'} \propto {\gamma'}^{-p}$ (for
$\gamma^{'}_{m}\leq\gamma^{'}\leq\gamma^{'}_{M}$) with the minimum
and maximum electron Lorentz factors ${\gamma'}_{m}
=\epsilon_{e}\frac{m_{p}(p-2)}{m_{e}(p-1)}(\Gamma-1)$ and
$\gamma'_M\sim q_eB'r/(m_ec^2)$, where the prime represents that the
quantities are measured in the comoving frame. The magnetic filed
strength $B'$ can be calculated with the parameter $\epsilon_B$. By
considering the radiative cooling of the electrons and introducing a
cooling Lorentz factor ${\gamma'}_{c} =6\pi
m_ec/(\sigma_{_{T}}{B'}^2\Gamma t)$, where $\sigma_{_{T}}$ is
Thomson cross section, a quasi-steady electron distribution can be
written as (Huang et al. 2000b)
\begin{equation}
{n_{{\gamma'}}^{'}=\frac{n{'}_{e}}{\gamma{'}_{l}}}\left\{
\begin{array}{ll}
\left(\frac{{\gamma'}}{{\gamma'}_{l}}\right)^{-x},
 &{\gamma'}_{l}\leq{\gamma'}\leq{\gamma'}_{h},\\
\left(\frac{{\gamma'}_{h}}{{\gamma'}_{l}}\right)^{-x}\left(\frac{{\gamma'}}{{\gamma'}_{h}}\right)^{-p-1},
 &{\gamma'}_{h}<{\gamma'}\leq \gamma'_{M}.
\end{array}\right.
\end{equation}
For ${\gamma'}_{c}>{\gamma'}_{m}$ (slow cooling), we have $x=p$,
$\gamma^{'}_{l}=\gamma^{'}_{m}$, and
$\gamma^{'}_{h}=\gamma^{'}_{c}$, whereas $x=2$,
$\gamma^{'}_{l}=\gamma^{'}_{c}$ and $\gamma^{'}_{h}=\gamma^{'}_{m}$
for ${\gamma'}_{m}>{\gamma'}_{c}$ (fast cooling). The comoving
synchrotron emission coefficient at frequency $\nu'$ can be given by
(Rybicki \& Lightman 1979)
\begin{eqnarray}
{j'}_{\nu'}=\frac{1}{4\pi}\frac{\sqrt{3}q_e^3B'}{m_ec^2}\int_{\gamma'_l}^{\gamma'_M}
{n'}_{{\gamma'}}\left[\frac{{\nu'}}{{\nu'}_0}
\int_{\nu^{'}\over{\nu'}_0}^{\infty}K_{5/3}(t)dt\right]d{\gamma'},
\end{eqnarray}
where ${\nu'}_0=3q_eB'{\gamma'}^2/(4\pi m_ec)$ and $K_{5/3}(t)$ is
the Bessel function. Then the synchrotron flux density received by
the observers can be calculated as
\begin{equation}
F_{\nu}(t)=\frac{(1+z)}{d_{L}^2}\int_0^{\theta_{j}}\frac{{j'}_{{\nu'}}V'}
{\Gamma^3(1-\beta\cos{\theta})^3}\frac{\sin{\theta}}{2}\cos{\theta}d\theta\label{fnu},
\end{equation}
where $\nu=\nu'/[\Gamma(1-\beta\cos \theta)]$ is the observational
frequency, $d_{L}$ is the luminosity distance of the GRB, and $V'$
is the comoving volume of emission region. Additionally, the
equal-arrival-time effect has also been taken into account in our
calculation.

\section{NUMERICAL RESULTS}

\begin{table}
\centering \caption{Parameters for the energy injections}
\begin{tabular}{c|ccc}
 \hline\hline
 the $i$th injection &\qquad
$t_{\rm start}(10^{4}\rm s)$\qquad \qquad & $t_{\rm end}(10^{4}\rm
s)$ & $Q (10^{46}\rm erg ~s^{-1})$\\\hline
 1&
$ 0.3 $\qquad & $ 1.4 $ & $ 15.0$
\\\hline
 2&
$2.0$\qquad & $7.3$ & $5.0$
\\\hline
3& $19.0$\qquad & $28.0$ & $1.1$
\\\hline
 4&
$55.0 $\qquad& $75.0$ & $1.2$
\\\hline
\end{tabular}
\end{table}

We use the model described above to reproduce the light curve of the
X-ray afterglow of GRB 050712. A comparison between the numerical
result with the observational data is presented in Figure 1. As
shown, the unusual tierd X-ray light curve requires four times of
independent energy injections into the external shock. The reduced
chi-square $\chi^{2}/\nu$ of the fitting is 1.253. The model
parameters are taken as follows. (1) GRB ejecta parameters: the
initial isotropic-equivalent kinetic energy
$E_{0}=2.5\times10^{51}\rm erg$, the initial Lorentz factor
$\Gamma_{0}=125$, and the half opening angle $\theta_{j}=0.1$; (2)
Microphysical parameters: $\epsilon_{B}=0.04$, $\epsilon_{e}=0.2$,
and $p=2.3$; (3) Circumstance Parameter: $A=3\times 10^{35}$, where
a wind-like circumstance is taken because a too high value of $p$ is
required in the ISM case which is inconsistent with the observed
spectral index; (4) Energy injection parameters: as listed in Table
1, where $q=0$ is adopted in all of the four injections; (5)
Redshift: $z=1$, where the average value of the Swift GRBs is taken
because there is no redshift measurement for GRB 050712. It should
be noted that the goodness of the fitting here is only judged by the
eye, but not by a precise statistical criterion, as usually did in
GRB afterglow modelings, so that the error bars of the parameters
can not be given. This because too many parameters are involved and
a more precise fitting is in fact not more informative. In any case
the main features of the light curve has been captured by our
fitting. Additionally, the three early flares are fitted by the
``Norris function" as $f(t)=
C\exp(-\tau_{1}/(t-t_{0})-(t-t_{0})/\tau_{2})$, where $C$ is the
intensity, $t_{0}$ is the beginning time, $\tau_{1}$ and $\tau_{2}$
are the rise and decay timescales, respectively (Norris et al.
2005). The obtained luminosity and integrated energy of the flares
can be regarded as a lower limit because the emission efficiencies
of the flares are unknown.
%Then
%we can get the luminosity distance of GRB 050712 as $D_{L}=6.7\rm
%~Gpc$, by using the $\Lambda$CDM ($\Lambda$ cold dark matter)
%concordance model of $\Omega _{M}=0.27 $, $\Omega _{\Lambda}=0.73$,
%and $H_{0}=71\rm km s^{-1} Mpc^{-1} $.

The three early flares and the following four energy injections
indicate that the central engine of GRB 050712 releases energy at
least seven times after the burst. The energy, duration, and
luminosity of these energy releases are given in Figure 2, where the
energy and luminosity during the flares are represented. Swenson \&
Roming (2013) gave the flares of GRB 050712 start time lower limit.
As it is seen, the injection of energy at different times is
comparable to each other on the order of magnitude of
$3\times10^{51}$ erg. The duration($t_{end,i}-t_{start,i}$) of the
energy released increases with time and thus the luminosity($Q_{i}$)
declines. These features could have important implications for the
properties of the central engine.

\section{Summary AND DISCUSSION}
\label{sect:discussion}

GRB 050712 is one of the most interesting Swift GRBs, whose X-ray
afterglow has four apparent shallow decay plateaus. In this paper,
we numerically calculated the dynamical evolution and the radiation
of the GRB external shock with multiple energy injections. The model
can explain the X-ray afterglow light curve of GRB 050712 well. This
indicates that the central engine of GRB 050712 intermediately
releases energy at least seven times after the burst, and the first
three ones produce X-ray flares directly via internal dissipation
and the latter four ones result in four plateaus by injecting energy
into the external shock. The energy released during these four
plateaus are similar, but the luminosity decreases significantly
with an increasing duration. Since the close connection between the
energy injections and the late activities of the central engine, our
result demonstrates that the GRBs with multiple energy injections
can be used to explore the properties of the GRB central engine. At
least, our result indicates that the GRB central engine can last
much longer than thought, which is related to the $t_{\rm burst}$
topic discussed in Zhang et al (2013). In addition, such a multiple
energy injection model may also have further implications for the
central-object-powered supernovae (Kasen \& Bildsten 2010) and
mergernovae (Yu et al. 2013).

For a single or a few plateaus, the plateau emission can in
principle be ascribed to a reverse shock propagating into the slow
tail of GRB ejecta (e.g., Uhm et al. 2012; Gao et al. 2013). The
tierd X-ray afterglow of GRB 050712 presented here unambiguously
break the degeneracy between the reverse shock model and the energy
injection model. Nevertheless, in this paper, we simply assumed that
the released energy is completely injected into the GRB external
shock. The energy released from the central engine is probably in
the form of Poynting flux initially, which could be accelerated and
thermalized before it catches up with the GRB external shock.
Therefore, the internal dissipation of the energy flow could
directly produces some afterglow emission such as the so-called
internal-origin afterglow (Yu et al. 2010; Zhang 2013). The internal
dissipation may arise from turbulent magnetic reconnection (Zhang \&
Yan 2011) or from the termination shock of the injected flow arising
from the collision between the injected flow and the GRB ejecta (Dai
2004; Yu \& Dai 2007; Yu et al. 2007; Mao et al. 2010; Dai \& Liu
2012).

\begin{acknowledgements}
This work made use of data supplied by the UK Swift Science Data
Centre at the University of Leicester. The authors thank the
instruction of Yun-Wei Yu who motivated this work. This work is
supported by the National Natural Science Foundation of China (Grant
No. 11103004), the Funding for the Authors of National Excellent
Doctoral Dissertations of China (Grant No. 201225), and the Program
for New Century Excellent Talents in University (Grant No.
NCET-13-0822).
\end{acknowledgements}

\label{lastpage}
\end{document}